\documentclass[twocolumn,prl,preprintnumbers,tightenlines,superscriptaddress,letterpaper]{revtex4}%footinbib,

\pdfoutput=1
\usepackage[english]{babel}
\usepackage{amsmath,amssymb,amsfonts,bm}
\usepackage{graphicx}
\usepackage{booktabs}
\usepackage{hyperref}
\usepackage{color}
\usepackage[sort&compress]{natbib}
\usepackage{tikz}
\usepackage{ytableau}

\usepackage{comment}

\usepackage{cleveref}
\newcommand*{\dk}[1]{\textcolor{red}{#1}}

\begin{document}

% =============================================================================
\title{Dynamics of Simplest Chiral Gauge Theories}
% =============================================================================

\author{Dan Kondo}
\email{dan.kondo@ipmu.jp}
\affiliation{Kavli Institute for the Physics and Mathematics of the
  Universe (WPI), University of Tokyo,
  Kashiwa 277-8583, Japan}

\author{Hitoshi Murayama}
\email{hitoshi@berkeley.edu, hitoshi.murayama@ipmu.jp, Hamamatsu Professor}
\affiliation{Leinweber Institute for Theoretical Physics, University of California, Berkeley, CA 94720, USA}
\affiliation{Kavli Institute for the Physics and Mathematics of the
  Universe (WPI), University of Tokyo,
  Kashiwa 277-8583, Japan}
\affiliation{Ernest Orlando Lawrence Berkeley National Laboratory, Berkeley, CA 94720, USA}

\author{Cameron Sylber}
\email{silvercam@berkeley.edu}
\affiliation{Department of Physics, University of California, Berkeley, CA 94720, USA}

\begin{abstract} 
Arguably, the simplest chiral gauge theories are $\mathrm{SO}(10)$ with $N_f$ fermion fields in the spinor representation {\bf 16}. We study their dynamics using their supersymmetric limits perturbed by an infinitesimal anomaly-mediated supersymmetry breaking as a guide. We predict the theory is gapped for $N_f=1,2$, while the $\mathrm{SU}(N_f)$ global symmetry is broken to $\mathrm{SO}(\mathrm{N}_f)$ for moderately large $N_f \geq 3$.
\end{abstract}

\maketitle

\paragraph{Introduction}

The fact that the weak interactions distinguish left from right came as a big surprise to physicists, as established by the $^{60}$Co $\beta$-decay experiment led by Chien-Shiung Wu \cite{Wu:1957my}. This idea of parity violation became the essential foundation of the Standard Model of particle physics, namely that it is based on {\it chiral}\/ gauge theories. Left-handed and right-handed particles have different quantum numbers, and actually they are fundamentally different particles unless the Higgs boson condensate connects them. 

Despite the essential importance of chiral gauge theories in describing nature, its definition has not been clear. Even though perturbation theory can be developed with somewhat ad hoc ways of regularizing divergences, a non-perturbative definition on the lattice has been difficult due to the fermion doubling problem \cite{Karsten:1980wd,Nielsen:1980rz,Nielsen:1981xu,Friedan:1982nk}. There has been a lot of progress in this area (see, {\it e.g.}\/, \cite{Kaplan:2009yg,Wang:2022ucy} for reviews and references therein), yet numerical simulation has not been possible for four-dimensional spacetime due to the sign problem. Therefore, it is of paramount importance to come up with other approaches in defining and working out chiral gauge theories.

Recently, one of the authors (HM) \cite{Murayama:2021xfj} proposed the methodology to use a supersymmetric (SUSY) version of gauge theories and perturb them by an infinitesimal supersymmetry breaking in the form called anomaly mediation \cite{Randall:1998uk,Giudice:1998xp}. In many cases, supersymmetry allows for exact understanding of non-perturbative effects in gauge theories, because the near-SUSY limit and non-SUSY limit appear to be continuously connected with no apparent sign of phase transition, and hence a cross over \cite{Kondo:2025njf}. In addition, anomaly mediation has the property called ``UV insensitivity'' such that its impact can be worked out whether the fields are elementary or composites. Therefore, the non-perturbative dynamics can be understood even in the presence of supersymmetry breaking. This methodology has already been applied to some chiral gauge theories \cite{Csaki:2021xhi,Csaki:2021aqv,Leedom:2025mcg,Goh:2025oes} with rather surprising results which did not agree with previous suggestions. 

In the end, we will need computer simulations to understand which suggestions correctly describe non-perturbative dynamics of chiral gauge theories. Kikukawa suggested that $\mathrm{SO}(10)$ gauge theories with fermions in spinor ({\bf 16}) representations are likely the first of such theories to be simulated on computers \cite{Kikukawa:2017ngf}. In fact, $\mathrm{SO}(10)$ is the smallest simple group that admits chiral fermions ({\it i.e.}\/, complex representations) and is anomaly free. Among them, the ${\bf 16}$-dimensional representation is the smallest complex representation. Kikukawa proposed a formulation for its lattice simulations \cite{Kikukawa:2017ngf} building on prior works \cite{Eichten:1985ft,Wen:2013ppa,Neuberger:1997fp,Ginsparg:1981bj,Kaplan:1992bt}.

In this Letter, we study dynamics of $\mathrm{SO}(10)$ gauge theories with $N_f$ Weyl fermions in the ${\bf 16}$ representation. Using the methodology of supersymmetry broken by anomaly mediation, we work out exact analytic solutions when the supersymmetry breaking is small. We also discuss what we might expect in the non-supersymmetric limit and find they are plausibly connected continuously. 

\paragraph{Supersymmetric $\mathrm{SO}(10)$ with {\bf 16}'s}

We can obtain non-perturbative superpotentials for the $\mathrm{SO}(10)$ theories with $N_f$ {\bf 16}s by first introducing additional $2(4-N_f)+1$ fields in the vector representation. Then the theories ``s-confine'' and the superpotentials are given already in \cite{Csaki:1996zb}. By adding mass terms to the extra fields and integrating them out, we obtain the desired superpotential of the original theories. Yet, the expressions in \cite{Csaki:1996zb} are somewhat sketchy without precise numerical coefficients and thus inadequate for our purpose. Below, we rederive the same results from the microscopic dynamics with precise expressions.

We consider the chiral superfields $\psi_\alpha$ $(\alpha=1, \cdots, N_f)$  that are in the ${\bf 16}$ representation of $\mathrm{SO}(10)$ (or $\mathrm{Spin}(10)$ to be more precise) and the fundamental representation of the $\mathrm{SU}(N_f)$ global symmetry. They have gauge-invariant polynomials $S_{\alpha\beta;\gamma\delta}$ constructed as
\begin{align}
	S_{\alpha\beta; \gamma\delta}
	&= \frac{1}{3!} A_{\alpha\beta}^{IJK} A_{\gamma\delta}^{IJK} = S_{\gamma\delta; \alpha\beta}, \nonumber\\
	A_{\alpha\beta}^{IJK}
	&= \psi_\alpha^T {\cal C}_{10} \Gamma^{IJK} \psi_\beta = -A_{\beta\alpha}^{IJK}. 
\end{align}
Here, $I,J,K=1,\cdots,10$ are $\mathrm{SO}(10)$ vector indices, and ${\cal C}_{10}$ is the ``charge conjugation'' matrix of $\mathrm{SO}(10)$ that satisfies ${\cal C}_{10} \Gamma^I {\cal C}_{10}^{-1} = - (\Gamma^{I})^T$. For our convention of $\mathrm{SO}(10)$ gamma matrices see the supplement.

We discuss cases $N_f = 1, 2, 3, 4$. For higher $N_f$, the theory has a magnetic description \cite{Berkooz:1997bb}. Unfortunately these do not have the $\mathrm{SU}(N_f)$ global symmetry manifest in the description. Here we do not attempt to study higher $N_f \geq 5$.

\paragraph{$N_f=1$ Case}

This theory was conjectured to dynamically break supersymmetry based on the difficulty in matching $U(1)_R$ anomalies \cite{Affleck:1984mf} and later supported by instanton analyses \cite{Amati:1988ft}. In either case, the theory is strongly coupled. It was proven later that the Witten index of this theory in fact vanishes \cite{Murayama:1995ng}. To place the theory under quantitative control, we introduce a chiral superfield in the vector representation of SO(10) with mass $M \ll \Lambda$. Without the AMSB effects, the theory spontaneously breaks both supersymmetry, leading to a massless goldstino, and $U(1)_R$, a massless Nambu--Goldstone boson. Once AMSB is included, both supersymmetry and $U(1)_R$ are explicitly broken and both of them acquire finite mass, making the theory gapped. See the supplemental material for details of the calculation.

Both supersymmetry and $\mathrm{U}(1)_R$ are explicitly broken with no global symmetry. Correspondingly, we confirmed that both the goldstino and the $\mathrm{U}(1)_R$ NGB acquire mass in the limit $m \ll M \ll \Lambda$ where a weakly-coupled analysis is justified, and hence the spectrum is indeed gapped.

\paragraph{$N_f=2$ Case}

The $D$-flat direction can be understood using the hierarchical symmetry breaking. Under the $\mathrm{SO}(8)$ subgroup, we have two pairs of $({\bf 8}_s\oplus{\bf 8}_a)$. Using the triality of $\mathrm{SO}(8)$, we can regard them as two $({\bf 8}_s\oplus{\bf 8}_v)$. While two ${\bf 8}_v$ breaks $\mathrm{SO}(10)$ to $\mathrm{SO}(8)$, one ${\bf 8}_s$ breaks it to $\mathrm{SO}(7)$ and then the other to $\mathrm{G}_2$. Recall that a spinor VEV of $\mathrm{SO}(7)$ breaks it to $\mathrm{G}_2$. 

Explicit $D$-flat direction for the classical moduli space is described by $S_{\alpha\beta; \gamma\delta}=S\epsilon_{\alpha\beta}\epsilon_{\gamma\delta}$ which is $\mathrm{SU}(2)$ singlet,
\begin{align}
	\psi_1 = \frac{v}{2} (\zeta_4 + \zeta_7) , &\quad
	\psi_2 = \frac{v}{2} (\zeta_1 + \zeta_6), 
\end{align}
giving $S = 3v^4/2$. See the supplemental material for the definitions of $\zeta_i$. In fact, $45-14=16\times 2-1$ and hence all $\psi_{1,2}$ are eaten except for the $D$-flat direction $S$.

The gaugino condensate of $\mathrm{G}_2$ generates the superpotential
\begin{align}
	W_{\it dyn} &= 4 \left( \frac{\Lambda^{20}}{S^2} \right)^{1/4} .
\end{align}

\begin{figure}[t]
\includegraphics[width=0.7\columnwidth]{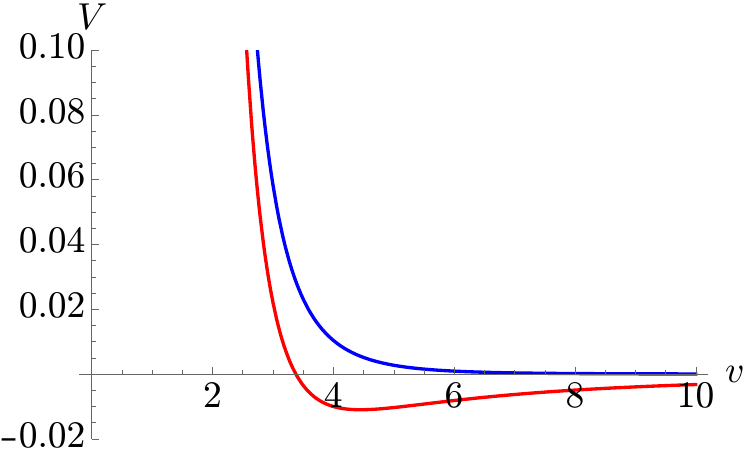}
  \caption{$N_f=2$ case, plot of the potential as a function of $v$ with $\Lambda=1$, without the anomaly mediation $m=0$ in blue, and with the anomaly mediation $m=0.01$ in red.}
  \label{fig:2flavor}
\end{figure}

With AMSB, the potential shown in~\cref{fig:2flavor} using the canonical K\"ahler potential along $v$ is
\begin{align}
	V &= \frac{128 \Lambda^{10}}{3v^6} -m\frac{80\Lambda^5}{\sqrt{6}\,v^2} ,
\end{align}
which settles to a minimum with no $\mathrm{SU}(2)$ breaking,
\begin{align}
	v &= \left(\frac{384}{25}\frac{\Lambda^{10}}{m^{2}}\right)^{1/8} . 
\end{align}
When $m\ll \Lambda$, the minimum is $v\gg \Lambda$ and hence the gauge coupling is weak, justifying the analysis using the canonical K\"ahler potential. The light spectrum consists of scalar of mass squared  $100m^2/3$, pseudo-scalar $50m^2/3$, and Majorana fermion $25m^2$, satisfying the vanishing supertrace. The spectrum is gapped.

\paragraph{$N_f=3$ Case}

The general $D$-flat direction breaks $\mathrm{SO}(10)$ to $\mathrm{SU}(2)$. It was non-trivial to come up with explicit parameterization of the general $D$-flat direction as is often the case for chiral gauge theories. It is helpful to observe that the unbroken $\mathrm{SU}(2)$ can be identified with $\mathrm{SU}(2)_{\text{R}}$ of the maximal Pati--Salam subgroup $\mathrm{SO}(10) \supset \mathrm{SO}(6) \times \mathrm{SO}(4) \simeq \mathrm{SU}(4) \times \mathrm{SU}(2)_{\text{L}} \times \mathrm{SU}(2)_{\text{R}}$ where the decomposition is
\begin{align}
	{\bf 16} &= ({\bf 4}, {\bf 2}, {\bf 1}) + ({\bf 4}^*, {\bf 1}, {\bf 2}).\nonumber
\end{align}
In order to keep $\mathrm{SU}(2)_{\text{R}}$ unbroken, the VEVs reside only in $({\bf 4}, {\bf 2}, {\bf 1})$. Out of 45 generators of $\mathrm{SO}(10)$, $15+3+3=21$ generators belong to the Pati--Salam group, while the remaining 24 generators are $({\bf 6}, {\bf 2}, {\bf 2})$ that map $({\bf 4}, {\bf 2}, {\bf 1})$ to $({\bf 4}^*, {\bf 1}, {\bf 2})$ and vice versa. As long as $({\bf 4}^*, {\bf 1}, {\bf 2})$ all vanish, $D$-flatness under $\mathrm{SU}(4) \times \mathrm{SU}(2)_{\text{L}}$ is sufficient to guarantee the $D$-flatness under the whole $\mathrm{SO}(10)$. Explicit $D$-flat direction for the classical moduli space is described by 
\begin{align}
    S_{\alpha\beta; \gamma\delta}
	&= \epsilon_{\alpha\beta\eta} \epsilon_{\gamma\delta\xi}
		\bar{S}^{\eta\xi}  
\end{align}
in the ${\bf 6^*}$ representation of the global $\mathrm{SU}(3)$ symmetry. We find the $D$-flat direction
\begin{align}
	\psi_1 &= v \left( \begin{array}{cc} 
		\cos\theta & 0 \\ 0 & 1 \\ 0 & 0 \\ 0 & 0
		\end{array} \right), \quad
	\psi_2 = v \left( \begin{array}{cc} 
		0 & 0 \\ 0 & 0 \\ 1 & 0 \\ 0 & \cos\theta
		\end{array} \right), \nonumber \\
	\psi_3 &= v \left( \begin{array}{cc} 
		\sin\theta\sin\phi & -\sin\theta\cos\phi \\ 0 & 0 \\ 0 & 0 \\ \sin\theta\cos\phi & \sin\theta\sin\phi
		\end{array} \right), \label{eq:Dflat3} 
\end{align}
which gives
\begin{align}
	\bar{S} &= v^4 \left( \begin{array}{ccc}
	0 & -12 \sin^2 \theta & 6\sin 2\theta \sin\phi\\
	-12 \sin^2 \theta & 0 & 6\sin 2\theta \sin\phi\\
	6\sin 2\theta \sin\phi & 6\sin 2\theta \sin\phi & -24\cos^2 \theta
	\end{array} \right). 
\end{align}
We verified that all possible eigenvalues can be generated by appropriate choice of $v, \theta, \phi$. 

\begin{figure}[t]
\includegraphics[width=0.7\columnwidth]{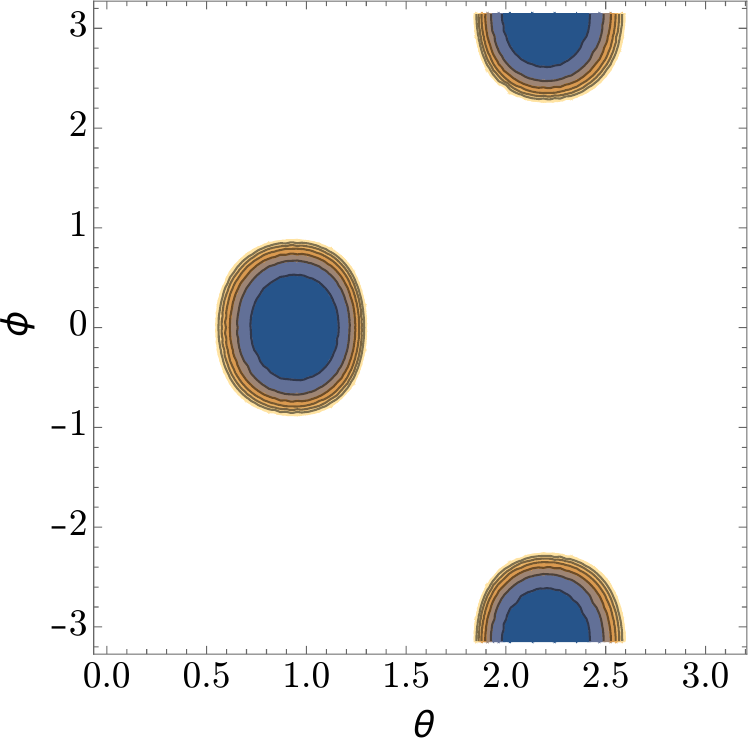}
\includegraphics[width=0.7\columnwidth]{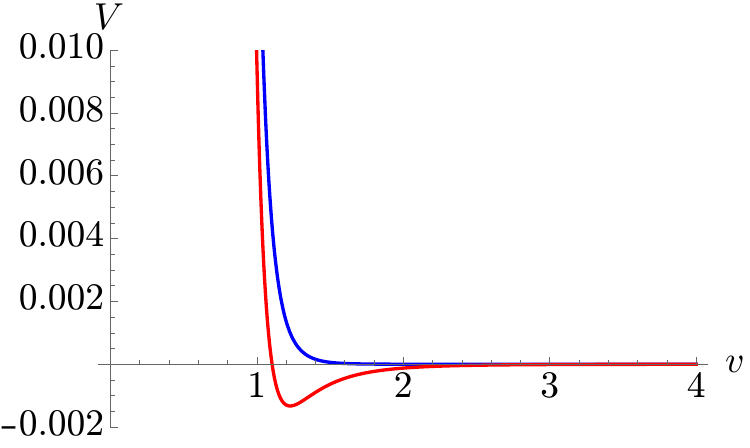}
  \caption{$N_f=3$ case, Top: Contour plot of the potential on the $(\theta,\phi)$ plane with $\Lambda=1$, $m=0.01$, and $v$ at its minimum Eq.~\eqref{eq:v3}. It shows the minimum at $\phi=0$ and $\theta=\arctan\sqrt{2} \approx 0.955$. The other minimum is equivalent under $(\theta,\phi)\simeq(-\theta,-\phi)$. Bottom: The behavior of the potential along the $v$ direction with (red) and without (blue) the anomaly mediation.}
  \label{fig:3flavor}
\end{figure}

Along the $D$-flat directions, $\mathrm{SO}(10)$ is generically broken to $\mathrm{SU}(2)_{\text{R}}$. Out of $16 \times 3=48$ chiral superfields, $45-3=42$ are eaten, leaving $48-42=6$ components described by $\bar{S}^{\eta\xi}$. The remaining $\mathrm{SU}(2)_{\text{R}}$ gauge group develops a gaugino condensate which leads to the run-away behavior of ${\rm det}\,\bar{S}$,
\begin{align}
	W_{\it dyn} &= 2 \left( \frac{\Lambda^{18}}{{\rm det}\,\bar{S}}\right)^{1/2} .
\end{align}
To work out the potential, it is important that all chiral superfields participate in its derivation even if they do not have an expectation value. We find
\begin{align}
	\lefteqn{
	V_{\rm SUSY} = \frac{\csc ^6(\theta ) \sec ^2(\theta ) \sec ^2(\phi ) \Lambda^{18}}{13824 v^{14}} }
	 \\
	&\times \left((\cos (2 \theta )+3) \sec^2(\theta ) \sec ^2(\phi )-9 \cos (2 \theta )+13\right),
	\nonumber
\end{align}
and
\begin{align}
	V_{\rm AMSB} = -\sqrt{\frac{3}{2}} \frac{ m \Lambda^{9}}{4 v^6}
   	\csc ^2(\theta ) \sec (\theta ) \sec (\phi ). %\nonumber
\end{align}
As shown in~\cref{fig:3flavor}, with AMSB, it settles to a minimum with $\theta=\arctan\sqrt{2}$ and $\phi=0$, where three eigenvalues of $\bar{S}$ are the same up to a unitary transformation. Therefore, the theory breaks 
$\mathrm{SU}(3)$ to $\mathrm{SO}(3)$. The minimum is at
\begin{align}
    v = \left( \frac{49\Lambda^{18}}{18432m^2} \right)^{1/16}
    \label{eq:v3}
\end{align}
and hence $v \gg \Lambda$ as long as $m \ll \Lambda$ which justifies the weakly-coupled analysis using the canonical K\"ahler potential. The light spectrum consists of multiplets ${\bf 6}^* = {\bf 1}+{\bf 5}$ under $\mathrm{SO}(3)$ with mass-squared
\begin{align}
    \mbox{scalars:}&\quad \frac{648}{7}m^2, \frac{162}{49}m^2 \times 5, \\
    \mbox{pseudo-scalars:}&\quad \frac{486}{7}m^2, 0 \times 5, \\
    \mbox{fermions:}&\quad 81m^2, \frac{81}{49}m^2 \times 5.
\end{align}
Five massless pseudo-scalars are the NGBs of $\mathrm{SU}(3)/\mathrm{SO}(3)$, and they together satisfy the vanishing supertrace.

\paragraph{$N_f=4$ Case}

In this case, the $\mathrm{SO}(10)$ gauge group is generically completely broken on the moduli space. Out of $16\times 4=64$ chiral superfields, $45$ are eaten, leaving $64-45=19$ moduli fields. They are described by ${\bf 20}$ of $\mathrm{SU}(4)$, or equivalently traceless symmetric tensor $S=S^T$, ${\rm Tr}\,S=0$ of $\mathrm{SO}(6)$ with one constraint. Unfortunately, the expression given in \cite{Csaki:1996zb} is too sketchy for our purpose. We find that the classical constraint is $4 {\rm Tr}\, S^4 - ({\rm Tr}\,S^2)^2=0$ by explicitly looking at many $D$-flat solutions. This factor of four is crucial because it tells us which symmetry breaking pattern is possible within the constraint. Quantum mechanically, the constraint is modified and we have the superpotential
\begin{align}
	W_{\rm dyn} = X \left( 4 {\rm Tr}\, S^4 - ({\rm Tr}\,S^2)^2 - \Lambda^{16} \right).
\end{align}

We studied the possible ground state. We list up the candidates in supplement.

As a result, We consider $\mathrm{SO}(6)/(\mathrm{SO}(3)\times \mathrm{SO}(3)) = \mathrm{SU}(4)/\mathrm{SO}(4)$ as the main candidate for the ground state. However this analysis is not rigorous, given that the theory is in the strongly coupled regime and we do not have control over the K\"ahler potential. Therefore we leave the possibility of $\mathrm{SO}(6)/\mathrm{SO}(5) = \mathrm{SU}(4)/\mathrm{Sp}(4)$ open to be conservative.

\paragraph{Non-Supersymmetric limits}

As $m$ is increased and approaches $\Lambda$, we cannot solve the theory any more. There may be a phase transition around $m \sim \Lambda$. If there isn't, the theory remains in the same universality class, namely that it has the same massless particle content and symmetry breaking pattern. We hope to learn non-SUSY limits this way. At the same time, it is important to discuss what we may expect in the non-SUSY limit to see if it is plausible that they are connected continuously. In this discussion, we abuse our notation to refer to Weyl fermions using the same symbol $\psi_\alpha$ as the chiral superfields above.

There is no fermionic composite operator, given that it would require an odd power of $\psi$, while each $\psi$ is odd under the $\mathbb{Z}_2$ center of $\mathrm{Spin}(10)$, and hence an odd power of $\psi$ cannot be gauge invariant. Therefore, there cannot be a massless composite fermion for the anomaly matching. The $\mathrm{SU}(N_f)$ global symmetry must be dynamically broken to an anomaly-free subgroup. The candidate subgroups are $\mathrm{SO}(N_f)$ or $\mathrm{Sp}(2[N_f/2])$. Using the Hilbert series techniques developed for the Standard Model Effective Field Theory (SMEFT) in \cite{Henning:2015alf}, one finds the number of gauge-invariant four-fermion operators to be $1, 6, 21, 56$ for $N_f=1,2,3,4$, respectively \footnote{We thank Risshin Okabe for providing this result to us.}. They can be written down as
\begin{align}
    Z_{\alpha\beta;\gamma\delta}
    &=  X_{\alpha\beta}^I X_{\gamma\delta}^I , \\
    X_{\alpha\beta}^I
    &= X_{\beta\alpha}^I
    \equiv \psi_{\alpha}^T C 
    {\cal C}_{10} \Gamma^I \psi_\beta,
\end{align}
where $C=i\gamma_0\gamma_2$ is the usual charge conjugation matrix of the $\mathrm{SO}(3,1)$ Lorentz group. 

In general, the quadrilinear condensate may be
\begin{align}
	Z_{\alpha\beta;\gamma\delta} 
	&\propto \delta_{\alpha\beta}\delta_{\gamma\delta}
	+ \delta_{\alpha\gamma}\delta_{\beta\delta}
	+ \delta_{\alpha\delta}\delta_{\beta\gamma}
	\label{eq:SONf}
\end{align}
to leave $\mathrm{SO}(N_f) \subset \mathrm{SU}(N_f)$ unbroken, or
\begin{align}
	Z_{\alpha\beta;\gamma\delta} 
	&\propto \delta_{\alpha\beta}\delta_{\gamma\delta}
	- \delta_{\alpha\delta}\delta_{\gamma\beta} \quad
	\mbox{or} \quad
	J_{\alpha\gamma}J_{\beta\delta} + J_{\alpha\delta}J_{\beta\gamma}
	\label{eq:SpNf}
\end{align}
to leave $SO(N_f)$ or $\mathrm{Sp}(2[N_f/2]) \subset \mathrm{SU}(N_f)$  unbroken. For the $N_f=4$ case, there is another singlet operator,
\begin{align}
    \epsilon^{\alpha\beta\gamma\delta}
    (\psi_\alpha^T C {\cal C}_{10} \Gamma^{IJK} \psi_\beta) 
    (\psi_\gamma^T C {\cal C}_{10} \Gamma^{IJK} \psi_\delta),
\end{align}
which is not relevant for symmetry breaking.

One of the few methods proposed to analyze dynamics of chiral gauge theory is called the tumbling \cite{Raby:1979my} (originally suggested in \cite{Georgi:1979md}). The tumbling hypothesis assumes a fermion bilinear condensate in the Most Attractive Channel (MAC), which suggests $SU(N_f)\rightarrow SO(N_f)$ for all $N_f$. See discussions in the supplemental material.

On the other hand, the analysis using supersymmetry above shows the preference for $\mathrm{SO}(N_f)$ for $N_f=3$ \eqref{eq:SONf}, while for $\mathrm{Sp}(2)=\mathrm{SU}(2)$ for $N_f=2$ \eqref{eq:SpNf}. For $N_f=4$ our result prefers $\mathrm{SO}(4)$ while it is inconclusive and may allow for $\mathrm{Sp}(4)$.

In particular, $N_f=2$ is special where the supersymmetric analysis suggests that the global $\mathrm{SU(2)}$ symmetry is unbroken. Namely the tumbling hypothesis and supersymmetric analysis differ in the predictions, which would be very interesting to be settled by future lattice simulations. It may be that $\mathrm{SU}(N_f)/\mathrm{Sp}(N_f)$ is the preferred ground state when $N_f$ is even. 

It is worthwhile recalling the results for other chiral gauge theories. \cite{Csaki:2021xhi,Csaki:2021aqv,Leedom:2025mcg,Goh:2025oes} It would be useful to study more examples to draw general lessons on chiral gauge theories.

\paragraph{Conclusion}

We studied dynamics of $\mathrm{SO}(10)$ gauge theory with $N_f$ fermions in the {\bf 16}-representation. This is arguably the simplest chiral gauge theory to be simulated on lattice in the near future. We solved the theory exactly with supersymmetry broken infinitesimally by anomaly mediation. The theory is gapped for $N_f=1,2$, while exhibits $\mathrm{SU}(3)/\mathrm{SO}(3)$ symmetry breaking for $N_f=3$. The case $N_f=4$ prefers $\mathrm{SU}(4)/\mathrm{SO}(4)$ while $\mathrm{SU}(4)/\mathrm{Sp}(4)$ possibility remains. Even though we cannot exclude a phase transition when supersymmetry breaking scale is increased, we find it quite plausible that the general pattern of symmetry breaking is $\mathrm{SU}(N_f)/\mathrm{SO}(N_f)$ for the non-supersymmetric limit for $N_f>2$. 

\paragraph{Acknowledgments}

\begin{acknowledgments}

We thank Risshin Okabe for listing up four-fermion operators using the Hilbert series for us. HM also thanks Yoshio Kikukawa for useful conversation and encouragement initiated at the YITP Workshop ``Strings and Fields 2021'' from August 23 to 27, 2021. DK and HM were supported by the Beyond AI Institute, the University of Tokyo. HM was also supported in part by the DOE under grant DE-AC02-05CH11231, by the NSF grant PHY-1915314, by the JSPS Grant-in-Aid for Scientific Research JP20K03942, MEXT Grant-in-Aid for Transformative Research Areas (A) JP20H05850, JP20A203, by WPI, MEXT, Japan, and Hamamatsu Photonics, K.K. 
\end{acknowledgments}

\section{Conventions}\label{sec:appendix}

We adopt the following conventions for the $\mathrm{SO}(10)$ gamma matrices. Using the standard Pauli matrices
\begin{align}
	\sigma_1 = \left( \begin{array}{cc} 0 & 1 \\ 1 & 0 \end{array} \right), &\quad
	\sigma_2 = \left( \begin{array}{cc} 0 & -i \\ i & 0 \end{array} \right), \nonumber \\
	\sigma_3 = \left( \begin{array}{cc} 1 & 0 \\ 0 & -1 \end{array} \right), &\quad
	\sigma_0 = \left( \begin{array}{cc} 1 & 0 \\ 0 & 1 \end{array} \right), 	
\end{align}
we define 32-by-32 gamma matrices that satisfy the Clifford algebra $\{\Gamma^I, \Gamma^J\} = 2 \delta^{IJ}$,
\begin{align}
\Gamma^1 &= \sigma_1 \otimes \sigma_0 \otimes \sigma_0 \otimes \sigma_0 \otimes \sigma_0, 
\nonumber \\
\Gamma^2 &= \sigma_2 \otimes \sigma_0 \otimes \sigma_0 \otimes \sigma_0 \otimes \sigma_0, 
\nonumber \\
\Gamma^3 &= \sigma_3 \otimes \sigma_1 \otimes \sigma_0 \otimes \sigma_0 \otimes \sigma_0, 
\nonumber \\
\Gamma^4 &= \sigma_3 \otimes \sigma_2 \otimes \sigma_0 \otimes \sigma_0 \otimes \sigma_0, 
\nonumber \\
\Gamma^5 &= \sigma_3 \otimes \sigma_3 \otimes \sigma_1 \otimes \sigma_0 \otimes \sigma_0, 
\nonumber \\
\Gamma^6 &= \sigma_3 \otimes \sigma_3 \otimes \sigma_2 \otimes \sigma_0 \otimes \sigma_0, 
\\
\Gamma^7 &= \sigma_3 \otimes \sigma_3 \otimes \sigma_3 \otimes \sigma_1 \otimes \sigma_0, 
\nonumber \\
\Gamma^8 &= \sigma_3 \otimes \sigma_3 \otimes \sigma_3 \otimes \sigma_2 \otimes \sigma_0, 
\nonumber \\
\Gamma^9 &= \sigma_3 \otimes \sigma_3 \otimes \sigma_3 \otimes \sigma_3 \otimes \sigma_1, 
\nonumber \\
\Gamma^{10} &= \sigma_3 \otimes \sigma_3 \otimes \sigma_3 \otimes \sigma_3 \otimes \sigma_2, 
\nonumber \\
\Gamma^{11} &= \sigma_3 \otimes \sigma_3 \otimes \sigma_3 \otimes \sigma_3 \otimes \sigma_3.
\nonumber
\end{align}
The {\bf 16} representation consists of the components $\Gamma^{11}=+1$. The generators of $\mathrm{SO}(10)$ are given by
\begin{align}
	\frac{1}{2} \Sigma^{IJ}, \quad \Sigma^{IJ}=\frac{i}{2} [ \Gamma^I, \Gamma^J ].
\end{align}
We also use the rank-three matrices 
\begin{align}
	\Gamma^{IJK} = \frac{i}{3!} \Gamma^{[I,} \Gamma^{J,} \Gamma^{K]},
\end{align}
where the indices $I,J,K$ are totally anti-symmetrized. The ``charge conjugation'' matrix is defined by
\begin{align}
	{\cal C}_{10} &= -\Gamma^2 \Gamma^4 \Gamma^6 \Gamma^8 \Gamma^{10}
	= \sigma_2 \otimes \sigma_1 \otimes \sigma_2 \otimes \sigma_1 \otimes \sigma_2,
\end{align}
which satisfies ${\cal C}_{10}=-{\cal C}_{10}^T = {\cal C}_{10}^{-1}$, and ${\cal C}_{10} \Gamma^I {\cal C}_{10}^{-1} = - (\Gamma^I)^T$.

Using this convention, the $D$-flat direction for the three flavor case Eq.(7) in the main text is in the $({\bf 4}, {\bf 2}, {\bf 1})$ components under the Pati--Salam subgroup $\mathrm{SU}(4) \times \mathrm{SU}(2)_L \times \mathrm{SU}(2)_R$, and hence we focus on them,
\begin{align}
	\psi &= v \left( \begin{array}{cc} 
	\zeta_1 & \zeta_5 \\ \zeta_2 & \zeta_6 \\ \zeta_3 & \zeta_7 \\ \zeta_4 & \zeta_8
	\end{array} \right),
\end{align}
where each of the components is given by
\begin{align}
	\zeta_1 
	= \uparrow \otimes \uparrow \otimes \uparrow \otimes \uparrow \otimes \uparrow,
	&\quad \zeta_5
	= \uparrow \otimes \uparrow \otimes \uparrow \otimes \downarrow \otimes \downarrow,
	\nonumber \\
	\zeta_2
	= \uparrow \otimes \downarrow \otimes \downarrow \otimes \uparrow \otimes \uparrow,
	&\quad \zeta_6
	= \uparrow \otimes \downarrow \otimes \downarrow \otimes \downarrow \otimes \downarrow,
	\nonumber \\
	\zeta_3
	= \downarrow \otimes \uparrow \otimes \downarrow \otimes \uparrow \otimes \uparrow, 
	&\quad \zeta_7
	= \downarrow \otimes \uparrow \otimes \downarrow \otimes \downarrow \otimes \downarrow,
	\nonumber \\
	\zeta_4
	= \downarrow \otimes \downarrow \otimes \uparrow \otimes \uparrow \otimes \uparrow, 
	&\quad \zeta_8
	= \downarrow \otimes \downarrow \otimes \uparrow \otimes \downarrow \otimes \downarrow.
\end{align}

 $S_{\alpha\beta;\gamma\delta}$ correspond to the following Young tableau of the $\mathrm{SU}(N_f)$ flavor symmetry,
\begin{align}
\ytableausetup{mathmode}
S_{\alpha\beta; \gamma\delta}=
\begin{ytableau}
\alpha & \gamma\\
\beta & \delta
\end{ytableau}
\end{align}
and for $N_f=3$,
\begin{align}
\ytableausetup{mathmode}
\bar{S}^{\eta\xi}=
\overline{
\begin{ytableau}
    \eta & \xi
    \end{ytableau}
    }
\end{align}

\section{Anomaly Mediation}

Anomaly mediation of supersymmetry breaking (AMSB) can be formulated with the Weyl compensator $\Phi = 1 + \theta^{2} m$ \cite{Pomarol:1999ie} that appears in the supersymmetric Lagrangian as
\begin{align}
	{\cal L} &= \int d^{4} \theta \Phi^{*} \Phi K + \int d^{2} \theta \Phi^{3} W + c.c.
\end{align}
Here, $K$ is the K\"ahler potential, $W$ is the superpotential of the theory, and $m$ is the size of supersymmetry breaking. When the theory is conformal, $\Phi$ can be removed from the theory by rescaling the fields $\phi_{i} \rightarrow \Phi^{-1} \phi_{i}$. On the other hand, violation of conformal invariance leads to supersymmetry breaking effects. Solving for auxiliary fields, the superpotential leads to the tree-level supersymmetry breaking terms
\begin{align}
	{\cal L}_{\rm tree} &= m \left( \phi_{i} \frac{\partial W}{\partial \phi_{i}} - 3 W \right)
	+ c.c.
	\label{eq:tree}
\end{align}
In addition, conformal invariance is anomalously broken due to the running of coupling constants, and there are loop-level supersymmetry breaking effects in the scalar masses and gaugino masses,
\begin{align}
	m_{i}^{2}(\mu) &= - \frac{1}{4} \dot{\gamma}_{i}(\mu) m^{2}, 
	\label{eq:m2} \\
	m_{\lambda}(\mu) &= - \frac{\beta(g^{2})}{2g^{2}}(\mu) m. \label{eq:mlambda}
\end{align}
Here, $\gamma_{i} = d \ln Z_{i}(\mu)/d\ln\mu$, $\dot{\gamma} = d\gamma_i/d\ln\mu$, and $\beta(g^{2}) = dg^2/d\ln\mu$. 

For asymptotically free gauge theories without a superpotential, $m_i^2$ is positive and the gaugino acquires a mass. Therefore the UV theory can correctly decouple scalars and gauginos. In the IR theory, we can apply the above formulae even for composite fields because they depend only on the particle content and interactions present at the respective energy scale. This is the property called ``UV insensitivity'' which allows us to study the IR behavior of the theory. In cases below, we rely on the tree-level AMSB Eq.~\eqref{eq:tree} which is justified in the weakly-coupled limits.

\section{$N_f=1$ Case}

This case does not have a $D$-flat direction where we can analyze the theory with large field amplitudes and hence weak coupling. It was called ``non-calculable'' by Affleck, Dine, and Seiberg \cite{Affleck:1984mf}, who made a plausibility argument that $\mathrm{U}(1)_R$ anomalies cannot be matched with a reasonable fermion content. On the other hand, the broken $\mathrm{U}(1)_R$ implied dynamical supersymmetry breaking (later firmed up by \cite{Nelson:1993nf}). Then the massless particle content must consist of one Nambu--Goldstone boson (NGB) of the broken $\mathrm{U}(1)_R$ and one goldstino of the broken supersymmetry. Together with AMSB, the $\mathrm{U}(1)_R$ symmetry as well as supersymmetry are explicitly broken. Therefore, both the goldstino and the $\mathrm{U}(1)_R$ NGB are massive and the theory should be gapped.

To verify this plausibility argument explicitly, we use a trick proposed by HM \cite{Murayama:1995ng} to include a chiral superfield $H({\bf 10})$ in the vector representation to make the theory ``calculable.'' In the end one introduces a mass term to 
$H$ as 
\begin{align}
	W_{\it tree} &= \frac{1}{2} M H^2
\end{align}
to the superpotential. Raising $M$ beyond $\Lambda$ does not lead to a phase transition thanks to the holomorphy \cite{Seiberg:1994bp}. The $D$-flat direction is parameterized by
\begin{align}
	H &= \frac{1}{\sqrt{2}} 
	(0, 0, 0, 0, 0, 0, 0, 0, i(H^+-H^-), (H^++H^-)), \\
	\psi &= \frac{1}{\sqrt{2}} (\uparrow \otimes \uparrow \otimes \uparrow \otimes \uparrow \otimes 
		\uparrow +  \downarrow \otimes \downarrow \otimes \downarrow \otimes 
		\downarrow \otimes \uparrow) \chi .
\end{align}
The $D$-flatness for the $(9,10)$ generator requires
\begin{align}
	\left| H^+ \right|^2 - \left| H^- \right|^2 - \frac{1}{2} \left| \chi \right|^2 = 0.
	\label{eq:Dflat}
\end{align}
Along this flat direction, $\mathrm{SO}(10)$ is broken to $\mathrm{SO}(7)$. It is obvious that $H$ above breaks $\mathrm{SO}(10)$ to $\mathrm{SO}(8)$, where ${\bf 16}$ decomposes as ${\bf 8}_s\oplus {\bf 8}_a$ as the spinor and anti-spinor. Using the triality of $\mathrm{SO}(8)$, the anti-spinor can be regarded as a vector. Its vacuum expectation value (VEV) breaks $\mathrm{SO}(8)$ further to $\mathrm{SO}(7)$ with a non-trivial embedding. Indeed, the Higgs mechanism in supersymmetry gives the counting $45-2\times 16=21-2$ where 2 classical moduli fields can be identified with $\psi\psi H$ and $H^2$. The gaugino condensate of $\mathrm{SO}(7)$ generates the dynamical superpotential
\begin{align}
	W_{\it dyn} &= 5 \left(\frac{\Lambda^{21}}{(\psi\psi H)^2} \right)^{1/5},\ 
	\psi\psi H = \sqrt{2} H^+ \chi^2.
\end{align}
It is easy to obtain a minimum that breaks supersymmetry with a massless goldstino and a massless $\mathrm{U}(1)_R$ NGB. The minimum obtained numerically is
\begin{align}
    H^+ &= 1.45 \left( \frac{\Lambda^{21}}{M^5} \right)^{1/16}, \\
    H^- &= 0.34 \left( \frac{\Lambda^{21}}{M^5} \right)^{1/16}, \\
    \chi &= 1.99 \left( \frac{\Lambda^{21}}{M^5} \right)^{1/16}.
\end{align}
As expected, the field values are much larger than $\Lambda$ when $M \ll \Lambda$ justifying the canonical K\"ahler potential used to find the minimum and the theory is perfectly calculable.

The Witten index vanishes because of the dynamically broken supersymmetry. Given that the Witten index is topologically invariant, it stays vanishing as $M \rightarrow \infty$, establishing the dynamical supersymmetry breaking in the $\mathrm{SO}(10)$ theory with a single ${\bf 16}$ \cite{Murayama:1995ng}.

\begin{figure}[t]
\includegraphics[width=0.7\columnwidth]{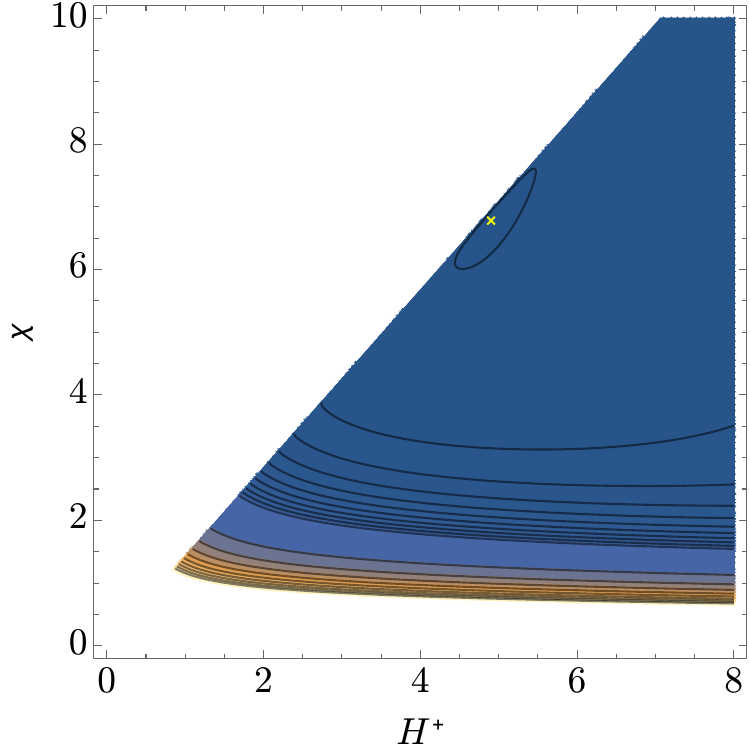}
  \caption{Contour plot of the potential on the $(H^+, \chi)$ plane with $M=0.01$, $m=0.0001$, $\Lambda=1$. The yellow cross is the location of the minimum. The region $\chi > \sqrt{2}H^+$ is cut off because it cannot satisfy the $D$-flatness condition Eq.~\eqref{eq:Dflat}.}
  \label{fig:1flavor}
\end{figure}

\section{Light fermion spectrum}
After finding the minimum, it is straghtfoward to calculate the mass spectrum. We make a mass matrix using the solution of $D$-flat direction.
\begin{align}
(M)_{ij}=\frac{\partial^2 W}{\partial\psi_i \partial\psi_j}|_{\psi_i,\psi_j:\text{minimizer}}
\end{align}

\subsection{$N_f=2$ Case}
With the superpotential
\begin{align}
 W_{\it dyn}
 =4\left(\frac{\Lambda^{20}}{S^2}\right)^{\frac{1}{4}}
 =4\sqrt{\frac{2}{3}}\frac{\Lambda^5}{v^2}
\end{align}
and the potential
\begin{align}
 V
 =\left(\frac{\partial W}{\partial v}\right)^2+2m\left(\frac{\partial W}{\partial v}v-3W\right)
 =\frac{128\Lambda^{10}}{3v^6}-\frac{80m\Lambda^5}{\sqrt{6}v^2}
\end{align}
The fermion mass is
\begin{align}
    \frac{\partial^2W}{\partial v^2}|_{v=v_{\it min}}
    =5m.
\end{align}
where 
\begin{align}
v_{\it min}=\left(\frac{384\Lambda^{10}}{25m^2}\right)^{\frac{1}{8}}.
\end{align}
is the minimum.
The square of the scalar mass is
\begin{align}
    \frac{1}{2}\frac{\partial^2V}{\partial v^2}|_{v=v_{\it min}}
    =\frac{100m^2}{3}.
\end{align}
For the pseudo scalar mass, we have to care about the imaginary part. To expand around $v_{\it min}$, we obtain the potential for the pseudo-scalar $a$
\begin{align}
V_p
&=\left|\frac{\partial W}{\partial v}\right|^2+m\left(\frac{\partial W}{\partial v}v-3W\right)+  c.c.|_{v=v_{\it min}+ia} \nonumber\\
&=-\frac{20}{3}\left(\frac{2}{3}\right)^{\frac{3}{4}}\sqrt{5}m^{\frac{3}{2}}\Lambda^{\frac{5}{2}}+\frac{25}{3}m^2a^2+\mathcal{O}(a^3)
\end{align}
We can obtain the square of the pseudo-scalar mass as $50m^2/3$. We can check the vanishing of supertrace 
\begin{align}
    \frac{100m^2}{3}+\frac{50m^2}{3}-2\cdot(5m)^2=0.
\end{align}

\subsection{$N_f=3$ Case}
$45-3=42$ fermions are eaten by the gauge multiplets. Remaining $16\times3-42=6$ are partners of non-zero VEVs. Once $\mathrm{SU(3)}$ is broken to $\mathrm{SO(3)}$, they form $6^\star=1+5$. For the obtained solution $(\theta,\phi)=(\arctan\sqrt{2},0)$, 
\begin{align}
\psi_1&=\alpha_1\bm{e}_1+\alpha_{16}\bm{e}_{16}\\
\psi_2&=\beta_{21}\bm{e}_{21}+\beta_{28}\bm{e}_{28}\\
\psi_3&=\delta_4\bm{e}_4+\delta_{25}\bm{e}_{25}\\
\alpha_1&=\frac{7^{1/8}}{2^{11/16}3^{5/8}m^{1/8}}, 
\alpha_{16}=\frac{7^{1/8}}{2^{11/16}3^{1/8}m^{1/8}}\nonumber\\
\beta_{21}&=\frac{7^{1/8}}{2^{11/16}3^{1/8}m^{1/8}}, 
\beta_{28}=\frac{7^{1/8}}{2^{11/16}3^{5/8}m^{1/8}}\nonumber\\
\delta_{4}&=\frac{-7^{1/8}}{2^{11/16}3^{5/8}m^{1/8}}, 
\delta_{25}=\frac{7^{1/8}}{2^{11/16}3^{1/8}m^{1/8}}\nonumber
\end{align}
Let us denote the nonzero set as \begin{align}
    N_0=\{\alpha_{1}, \alpha_{16}, \beta_{21}, \beta_{28}, \delta_{4}, \delta_{25}\}.
    \end{align}
    The superpotential is
\begin{align}
    W_{\it dyn}
    =2\left(\frac{\Lambda^{18}}{{\rm det}\,\bar{S}}\right)^{\frac{1}{2}}
    =\frac{1}{24\sqrt{6}}\frac{1}{\sqrt{\alpha_1\alpha_{16}^3\beta_{21}^3\beta_{28}\delta_{4}^2\delta_{25}^2}}.
\end{align}
The fermion mass matrix is
\begin{align}
    \frac{\partial^2 W_{\it dyn}}{\partial \psi_i \partial \psi_j}|_{N_0}
    =\frac{9m}{14}
    \begin{pmatrix}
    3&\sqrt{3}&\sqrt{3}&1&-\sqrt{2}&\sqrt{2}\\
    \sqrt{3}&5&3&\sqrt{3}&-\sqrt{6}&\sqrt{6}\\
    \sqrt{3}&3&5&\sqrt{3}&-\sqrt{6}&\sqrt{6}\\
    1&\sqrt{3}&\sqrt{3}&3&-\sqrt{2}&\sqrt{2}\\
    -\sqrt{2}&-\sqrt{6}&-\sqrt{6}&-\sqrt{2}&4&-2\\
    \sqrt{2}&\sqrt{6}&\sqrt{6}&\sqrt{2}&-2&4
    \end{pmatrix}.
\end{align}
The eigenvalues of this matrix (fermion mass) are \begin{align}
\left(9m, \frac{9}{7}m, \frac{9}{7}m, \frac{9}{7}m, \frac{9}{7}m, \frac{9}{7}m\right),
\end{align} one singlet and one quintet as expected. 
The potential is
\begin{align}
    V=\sum_{n_0\in N_0}\left(\frac{\partial W_{\it dyn}}{\partial n_0}\right)^2-2m\left(\sum_{n_0\in N_0}n_0\frac{\partial W_{\it dyn}}{\partial n_0}-3W\right)
\end{align}
The matrix of the scalar mass square is
\begin{align}
\frac{81m^2}{98}
    \begin{pmatrix}
13& 9\sqrt{3}& 9\sqrt{3}& 9& -9 \sqrt{2}& 9 \sqrt{2}\\
9 \sqrt{3}& 31& 27& 9 \sqrt{3}& -9\sqrt{6}& 9 \sqrt{6}\\
9\sqrt{3}& 27& 31& 9\sqrt{3}& -9\sqrt{6}& 9\sqrt{6}\\
9& 9\sqrt{3}& 9\sqrt{3}& 13& -9\sqrt{2}& 9\sqrt{2}\\
-9\sqrt{2}& -9\sqrt{6}& -9\sqrt{6}& -9\sqrt{2}& 22& -18\\
9\sqrt{2}& 9\sqrt{6}& 9\sqrt{6}& 9\sqrt{2}& -18& 22
    \end{pmatrix}.
\end{align}
The eigenvalues of this matrix (scalar mass squared) are
\begin{align}
    \left(\frac{648}{7}m^2, \frac{162}{49}m^2, \frac{162}{49}m^2, \frac{162}{49}m^2, \frac{162}{49}m^2, \frac{162}{49}m^2\right).
\end{align}
For pseudo-scalar case, expand around the vev caring about the imaginary part
\begin{align}
\psi_1&=(\alpha_1+ia_1)\bm{e}_1+(\alpha_{16}+ia_{16})\bm{e}_{16}\\
\psi_2&=(\beta_{21}+ib_{21})\bm{e}_{21}+(\beta_{28}+ib_{28})\bm{e}_{28}\\
\psi_3&=(\delta_4+id_4)\bm{e}_4+(\delta_{25}+id_{25})\bm{e}_{25}.
\end{align}
By taking the quadratic part of
\begin{align}
    N_{\Im}=\{a_{1}, a_{16}, b_{21}, b_{28}, d_{4}, d_{25}\},
\end{align}
we can obtain the square mass matrix of the pseudo-scalar
\begin{align}
\frac{81m^2}{14}
    \begin{pmatrix}
1& \sqrt{3}& \sqrt{3}& 1& -\sqrt{2}& \sqrt{2}\\
\sqrt{3}& 3& 3& \sqrt{3}& -\sqrt{6}& \sqrt{6}\\ 
\sqrt{3}& 3& 3& \sqrt{3}& -\sqrt{6}& \sqrt{6}\\
1& \sqrt{3}& \sqrt{3}& 1& -\sqrt{2}& \sqrt{2}\\
-\sqrt{2}& -\sqrt{6}& -\sqrt{6}& -\sqrt{2}& 2& -2\\
\sqrt{2}& \sqrt{6}& \sqrt{6}& \sqrt{2}& -2& 2\\
    \end{pmatrix}.
\end{align}
The eigenvalues of this matrix (pseudo-scalar mass squared) are
\begin{align}
    \left(\frac{486}{7}m^2, 0, 0, 0, 0, 0\right),
\end{align}
there are $\mathrm{SU}(3)/\mathrm{SO}(3)$ ($8-3=$) 5 Nambu-Goldstone bosons as expected.
 From the spectrum, we can check the vanishing of the supertrace 
 \begin{align}
     &\left\{\frac{648}{7}+5\times \frac{162}{49}\right\}m^2+\frac{486}{7}m^2-2\left\{9^2+5\times\left(\frac{9}{7}\right)^2\right\}m^2\nonumber\\
     &=0
 \end{align}
 .

\section{generality of the parametrization of $N_f=3$ case}
We show here that the following configuration
\begin{align}
	\psi_1 &= v \left( \begin{array}{cc} 
		\cos\theta & 0 \\ 0 & 1 \\ 0 & 0 \\ 0 & 0
		\end{array} \right), \quad
	\psi_2 = v \left( \begin{array}{cc} 
		0 & 0 \\ 0 & 0 \\ 1 & 0 \\ 0 & \cos\theta
		\end{array} \right), \nonumber \\
	\psi_3 &= v \left( \begin{array}{cc} 
		\sin\theta\sin\phi & -\sin\theta\cos\phi \\ 0 & 0 \\ 0 & 0 \\ \sin\theta\cos\phi & \sin\theta\sin\phi
		\end{array} \right), \label{eq:Dflat3} 
\end{align}
which gives
\begin{align}
	\bar{S} &= v^4 \left( \begin{array}{ccc}
	0 & -12 \sin^2 \theta & 6\sin 2\theta \sin\phi\\
	-12 \sin^2 \theta & 0 & 6\sin 2\theta \sin\phi\\
	6\sin 2\theta \sin\phi & 6\sin 2\theta \sin\phi & -24\cos^2 \theta
	\end{array} \right),
\end{align}
is the general parameterization of the $D$-flatness condition. First, we will check that this is the $D$-flat configuration. We can check that
\begin{align}
&\psi_1\Sigma_{ij}\psi_1 \nonumber\\
=&
-\begin{pmatrix}
    -(1+\cos^2\theta)J\\
    & \sin^2\theta J& & & \\
    & & \sin^2\theta J& & \\
    & & & \sin^2\theta J& \\
    & & & & \sin^2\theta J
\end{pmatrix}\\
&\psi_2\Sigma_{ij}\psi_2 \nonumber\\
=&
\begin{pmatrix}
    -(1+\cos^2\theta)J\\
    & \sin^2\theta J& & & \\
    & & -\sin^2\theta J& & \\
    & & & \sin^2\theta J& \\
    & & & & \sin^2\theta J
\end{pmatrix}\\
&\psi_3\Sigma_{ij}\psi_3 
=
\begin{pmatrix}
    \text{\huge{0}}_4\\
    & 2\sin^2\theta J&  \\
    & & \text{\huge{0}}_4
\end{pmatrix}\\
J&=
\begin{pmatrix}
    0&-1\\
    1&0
\end{pmatrix}.
\end{align}
To sum them up, we can explicitly check that they give the $D$-flat configuration.
\begin{align}
    \psi_1\Sigma_{ij}\psi_1+\psi_2\Sigma_{ij}\psi_2+\psi_3\Sigma_{ij}\psi_3=0
\end{align}
Second, we will check the generality. As written in the main text, there are six degrees of freedom. Among them, three out of six can be removed by the flavor symmetry. Therefore, the number of faithful parameters is three. There is one to one correspondence between the eigenvalue and the degree of freedom. From the expression of $S$, we can explicitly check that the number of parameters is three $(v,\theta,\phi)$ and three eigenvalues are 
\begin{align}
    \left\{6(1-\cos2\theta),-3(3+\cos2\theta+K),-3(3+\cos2\theta-K)\right\}\nonumber\\
K=\sqrt{5+\cos2\theta(6+5\cos2\theta)-4\cos2\phi\sin^22\theta}
\end{align}
(can be the same eigenvalues depending on the parameter values)

\section{$N_f=4$ Case}

Quantum mechanically, the constraint is modified and we have the superpotential
\begin{align}
	W_{\rm dyn} = X \left( 4 {\rm Tr}\, S^4 - ({\rm Tr}\,S^2)^2 - \Lambda^{16} \right).
\end{align}
The minimum turns out to be near $\Lambda$ and is not weakly coupled. Yet the fact that anomaly matching conditions are satisfied using $S$ as the dynamical degree of freedom implies that the K\"ahler potential is non-singular for $S$. We use a canonical K\"ahler potential for the $S$ field as well as the Lagrange multiplier field $X$ for the analysis. 

We studied following possible ground states,
\begin{align}
	\mathrm{SO}(4)\times \mathrm{SO}(2): &\quad S \propto {\rm diag}(1,1,1,1,-2,-2), \\
	\mathrm{SO}(2)^3: &\quad S \propto {\rm diag}(a,a,b,b,-a-b,-a-b), \\
	\mathrm{SO}(3)\times \mathrm{SO}(3): &\quad S\propto {\rm diag}(1,1,1,-1,-1,-1), \\
	\mathrm{SO}(5): &\quad S\propto {\rm diag}(1,1,1,1,1,-5), \\
	\mathrm{SO}(4): &\quad S\propto {\rm diag}(a,a,a,a,b,-4a-b).
\end{align}
The first two satisfy the classical constraint but not the quantum modified constraint, and hence are not on the moduli space quantum mechanically. The last one is minimized for $b=-5a$ which is actually the $\mathrm{SO}(5)$ configuration. Its vacuum energy is $V_0 = -0.000052\Lambda^4$ for $m=0.01\Lambda$, but has a tachyonic pseudo-scalar in the traceless symmetric tensor direction of $\mathrm{SO}(5)$ and it is hence unstable. The $\mathrm{SO}(3)\times \mathrm{SO}(3)$ configuration has the vacuum energy $V_0 = -0.00017\Lambda^4$ and is the deepest. Note also that there is no run-away direction given that $m^2_\psi>0$ \eqref{eq:m2} in the UV.

\section{Tumbling hypothesis}

We could not find applications of the tumbling hypothesis to the $\mathrm{SO}(10)$ theory with {\bf 16}s in the literature, and we \dk{will} attempt it here. Between two ${\bf 16}_1$ and ${\bf 16}_2$, there are ${\bf 16}^2 = {\bf 10}_S \oplus {\bf 120}_A \oplus {\bf 126}_S$ channels. The one-gauge-boson-exchange potential is given by 
\begin{align}
    V = \frac{g^2}{r} \sum_a T^a_{1}  T^a_2 = \frac{g^2}{r} \frac{1}{2} (C_{{\bf 16}^2}-2 C_{\bf 16})
\end{align}
with the quadratic Casimir operators $C_{\bf R}$, and we find
\begin{align}
	\sum_a T^a_{1}  T^a_2 &= \left\{ \begin{array}{ccc}
		-27/8 & & {\bf 10} \\
		-3/8 & & {\bf 120} \\
		+5/8 & & {\bf 126}
		\end{array} \right. .
\end{align}
The MAC then suggests
\begin{equation}
	\langle \psi_\alpha^T {\cal C}_{10} \Gamma^I \psi_\beta \rangle
	= \langle \psi_\beta^T {\cal C}_{10} \Gamma^I \psi_\alpha \rangle
	\propto \Lambda^3 \delta^{I,10} \delta_{\alpha\beta},
	\label{eq:tumbling}
\end{equation}
which breaks $\mathrm{SO}(10)$ gauge group to $\mathrm{SO(9)}$, while the global symmetry $\mathrm{SU}(N_f)$ to $\mathrm{SO}(N_f)$. The fermions $\psi_\alpha ({\bf 16})$ are real representations under the $\mathrm{SO(9)}$ gauge group and hence the theory is vector-like. It allows for the standard chiral symmetry breaking consistent with the remaining $\mathrm{SO}(N_f)$ symmetry, resulting in the confined massive Majorana fermions. However, an order parameter such as \eqref{eq:tumbling} cannot be taken at face value because gauge-non-invariant operators cannot have VEVs \cite{Elitzur:1975im}. This argument can be taken only as suggestive at best.

\bibliographystyle{utcaps_mod}
\bibliography{chiralrefs}

\end{document}